\documentclass[aps,prb,twocolumn,shortbibliography,superscriptaddress,article]{revtex4-1}
\usepackage{epsfig}
\usepackage{epstopdf}
\usepackage{amsmath}
\usepackage{amsfonts}
\usepackage{amssymb}
\usepackage{hyperref}
\usepackage{bm}
\usepackage{makecell}
\usepackage{rotating}
\usepackage{hyperref}
\usepackage{multirow}
\usepackage{graphicx}

\usepackage{graphicx}
\usepackage{dcolumn}
\usepackage{bm}
\usepackage{color}

\usepackage{tikz,xcolor,hyperref}

\definecolor{lime}{HTML}{A6CE39}
\DeclareRobustCommand{\orcidicon}{%
	\begin{tikzpicture}
	\draw[lime, fill=lime] (0,0)
	circle [radius=0.16]
	node[white] {{\fontfamily{qag}\selectfont \tiny ID}};
	\draw[white, fill=white] (-0.0625,0.095)
	circle [radius=0.007];
	\end{tikzpicture}
	\hspace{-2mm}
}

\foreach \x in {A, ..., Z}{%
	\expandafter\xdef\csname orcid\x\endcsname{\noexpand\href{https://orcid.org/\csname orcidauthor\x\endcsname}{\noexpand\orcidicon}}
}

\begin{document}


\title{Relativistic Spin-momentum locking in altermagnets}

\author{Carmine Autieri\orcidB}
\affiliation{International Research Centre Magtop, Institute of Physics, Polish Academy of Sciences,
Aleja Lotnik\'ow 32/46, PL-02668 Warsaw, Poland}
\affiliation{SPIN-CNR, UOS Salerno, IT-84084 Fisciano (SA), Italy}

\author{Amar Fakhredine\orcidF}
\affiliation{Institute of Physics, Polish Academy of Sciences, Aleja Lotnik\'ow 32/46, 02668 Warsaw, Poland}

\date{\today}

\begin{abstract}
Spin–momentum locking in altermagnets has been deeply explored in the non-relativistic limit. Including spin–orbit coupling, altermagnets exhibit antisymmetric exchange interactions, leading to spin cantings. Therefore, the spin-momentum locking differs among the three spin components S$_x$, S$_y$, and S$_z$, forming the relativistic spin-momentum locking. We consider orthorhombic YVO$_3$ and hexagonal MnTe. For YVO$_3$, the relativistic locking comprises s-, d$_{xy}$-, and d$_{xz}$-wave. In MnTe, the dominant component S$_y$ of MnTe inherits the polarized charge distribution and the non-relativistic spin-momentum locking bulk g-wave, but the breaking of the C$_{6z}$ rotational symmetry by the N\'eel vector lowers the symmetry from g-wave to d-wave. The relativistic spin-momentum locking for MnTe is composed of d$_{xz}$-, d$_{yz}$- and s-wave. Despite small magnitudes in real space, the canted spin components contribute significant spectral weight in k-space, impacting k-space properties such as the spin-Hall conductivity.
\end{abstract}


\maketitle


One of the most significant recent breakthroughs in the field of magnetism was the discovery of spin–momentum locking in the non-relativistic limit for magnetic systems with zero net magnetization.\cite{Smejkal22beyond} These conditions occur when the electronic states of spin-up and spin-down atoms are mapped onto each other not through translation or inversion, as in conventional antiferromagnets, but instead via rotational symmetries—whether proper or improper, symmorphic or nonsymmorphic.\cite{hayami2019momentum,hayami2020bottom,Smejkal22,Smejkal22beyond,Yuan2023,yuan2023degeneracy} Materials exhibiting this property were named altermagnets, and the discovery of spin–momentum locking in altermagnets has represented a true paradigm shift, revolutionizing the field of magnetism.
The spin-momentum locking of altermagnets can be used for several applications, as the magneto-tunneling junction\cite{PhysRevX.12.011028,Samanta2025}, Magneto-Optical Effects\cite{Sun2025}, piezomagnetism\cite{Xu2025}, elasto-Hall conductivity\cite{PhysRevB.111.184408}, non-linear Hall effect\cite{PhysRevLett.133.106701,han2025discoverylargemagneticnonlinear}, interplay with superconductivity\cite{cv8s-tk4c,PhysRevLett.131.076003} and several others\cite{Song2025,mazin2022altermagnetism}.

As a consequence of their rotational symmetries, altermagnets host antisymmetric exchange interactions in the presence of spin–orbit coupling (SOC). The simplest example of antisymmetric exchange is the staggered Dzyaloshinskii–Moriya interaction (DMI), which gives rise to relativistic weak ferromagnetism in altermagnets. Importantly, SOC itself preserves time-reversal symmetry and, in systems with Kramers degeneracy, cannot generate net magnetization. In altermagnetic compounds, however, time-reversal symmetry is already broken, and SOC can therefore induce the so-called weak ferromagnetism\cite{DZYALOSHINSKY1958241}. The altermagnetism is thus a necessary condition for the emergence of weak ferromagnetism.
Studies of antisymmetric exchange have primarily focused on weak ferromagnetism, corresponding to s-wave magnetism\cite{PhysRevLett.132.176702,839n-rckn,PhysRevLett.134.196703} and how the antisymmetric exchange evolves when the inversion is broken\cite{PhysRevLett.132.226702}. Yet the same antisymmetric exchange can also cant spins without producing a net moment. While spin canting has traditionally been discussed only in the context of weak ferromagnetism, canted configurations with zero net magnetization are usually overlooked. Here we demonstrate that such canted components can host their own spin–momentum locking and contribute a significant spectral weight to the band structure. The spin splitting arising from spin–orbit coupling was analyzed within the framework of multipole expansions. In the presence of spin–orbit coupling, magnetic states can be systematically classified according to their magnetic point groups, with the corresponding magnetic multipoles for each symmetry group being well established.\cite{Watanabe2018,Hayami2018,Yatsushiro2021} Nevertheless, to the best of our knowledge, no prior investigations have addressed these phenomena within a first-principles computational framework.

\begin{figure*}[t!]
\centering
\includegraphics[width=16.24cm,angle=0]{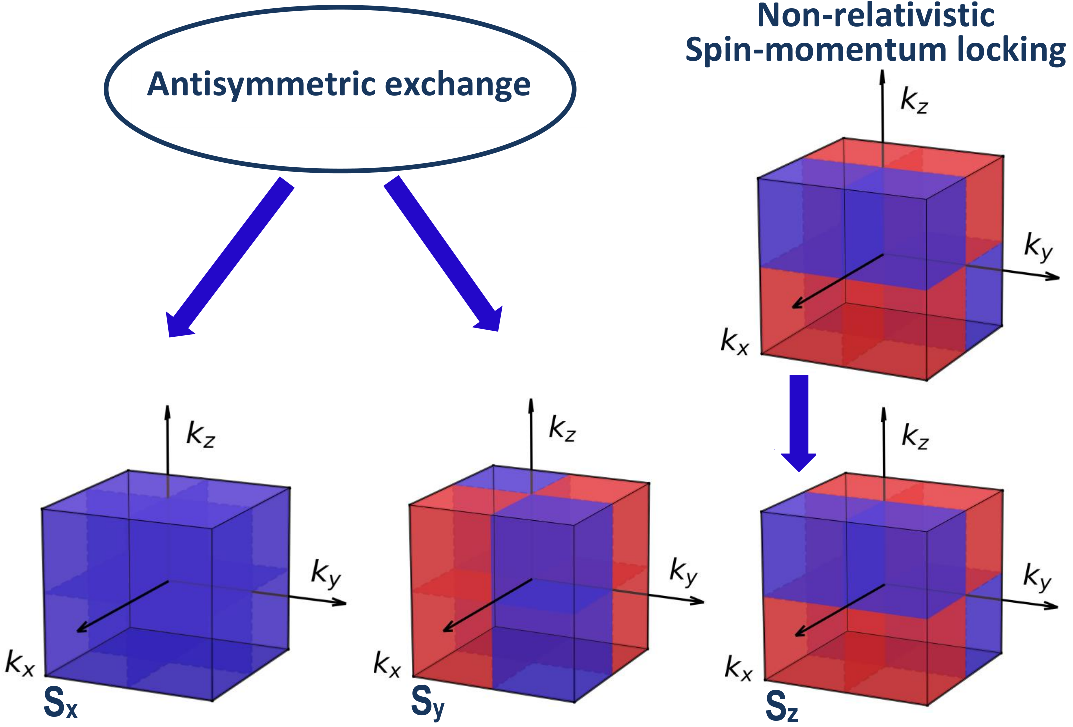}
\caption{Non-relativistic d-wave spin-momentum locking of G-type YVO$_3$ in the top part. Relativistic Spin-momentum locking for G-type YVO$_3$ with N\'eel vector along the $z$-axis in the bottom part. The S$_z$ component is the dominant component and inherits the d$_{xz}$-wave spin-momentum locking from the non-relativistic case. The S$_x$ component is s-wave due to the weak ferromagnetism, while the S$_y$ component is a d$_{xy}$-wave. Red and blue represent regions of the Brillouin zone with opposite spin-splitting. The weak ferromagnetism is represented by the s-wave with a complete Brillouin zone.}\label{RSML_YVO_Neelz}
\end{figure*} 


In this Letter, we aim to extend the concept of the non-relativistic spin-moment locking in altermagnets to the relativistic case using density functional theory (DFT).
We will study the relativistic spin-momentum locking for all three spin components S$_x$, S$_y$, and S$_z$, unveiling its properties and limitations. Using the orthorhombic YVO$_3$ and the hexagonal MnTe as representative altermagnetic compounds, we will show how the spin-momentum locking evolves in the relativistic limit. For YVO$_3$, we will focus on the G-type magnetic order, which is the ground state at low temperature\cite{PhysRevB.69.075110,Benckiser2008}. The crystal structures and the magnetic orderings of the two compounds are reported in Fig. S1 of the supplementary materials.



The non-relativistic spin-momentum locking of YVO$_3$ is a bulk d-wave with nodal planes for k$_x$=0 and k$_z$=0 and can be identified as d$_{xz}$\cite{Cuono23orbital,daghofer2025altermagneticpolarons}. If we consider YVO$_3$ with N\'eel vector along the $z$-axis, the S$_z$ component is the dominant component and it inherits the non-relativistic spin-momentum locking. The two subdominant spin components are obtained from the antisymmetric exchange, which produces the spin canting. In YVO$_3$, the antisymmetric exchange is provided by the staggered DMI\cite{autieri2024staggereddzyaloshinskiimoriyainducingweak}, given by the formula $\bold{D_{ij}\cdot[S_i \times S_j]}$ and such that the sum of the DMI vectors $\bold{D_{ij}}$ is zero, since the system is centrosymmetric. The DMI vectors for the perovskites with space group 62 can be located on the oxygen atoms between the magnetic V atoms \cite{PhysRevB.86.094413}. A toy model to show how the staggered DMI produces weak ferromagnetism is reported in Fig. S2 in the Supplemental Material. The results for the size and symmetries of the spin components are summarized in Table S1, and the spin-resolved band structure is reported in Fig. S3. The S$_x$ component exhibits weak ferromagnetism, while the S$_y$ component exhibits a magnetic ordering named C-type. The C-type magnetic order produces a non-relativistic spin-momentum locking, which is a bulk d-wave with nodal planes for k$_x$=0 and k$_y$=0 and we can define it as d$_{xy}$. The S$_y$ component exhibits spin-momentum locking, which is a d$_{xy}$-wave, coherent with the non-relativistic spin-momentum locking of the C-type.
In Fig. \ref{RSML_YVO_Neelz}, we report a schematic figure explaining how the non-relativistic spin-momentum and the antisymmetric exchange produce the relativistic spin-momentum locking of YVO$_3$ when the N\'eel vector is along the $z$-axis. 
The relativistic spin-momentum locking of YVO$_3$ with N\'eel vector along the $z$-axis is composed of s-wave for the S$_x$ component, d$_{xy}$-wave for the S$_y$ and d$_{xz}$-wave for the S$_z$ components. 
If the N\'eel vector were pointing along a different direction, the antisymmetric exchange interaction would produce a different spin canting (see Table I of supplementary materials) and the relativistic spin-momentum locking would be different (see Figs. S4 and S5 of the supplementary materials). 

\begin{figure}[t!]
\centering
\includegraphics[width=2.99cm,angle=270]{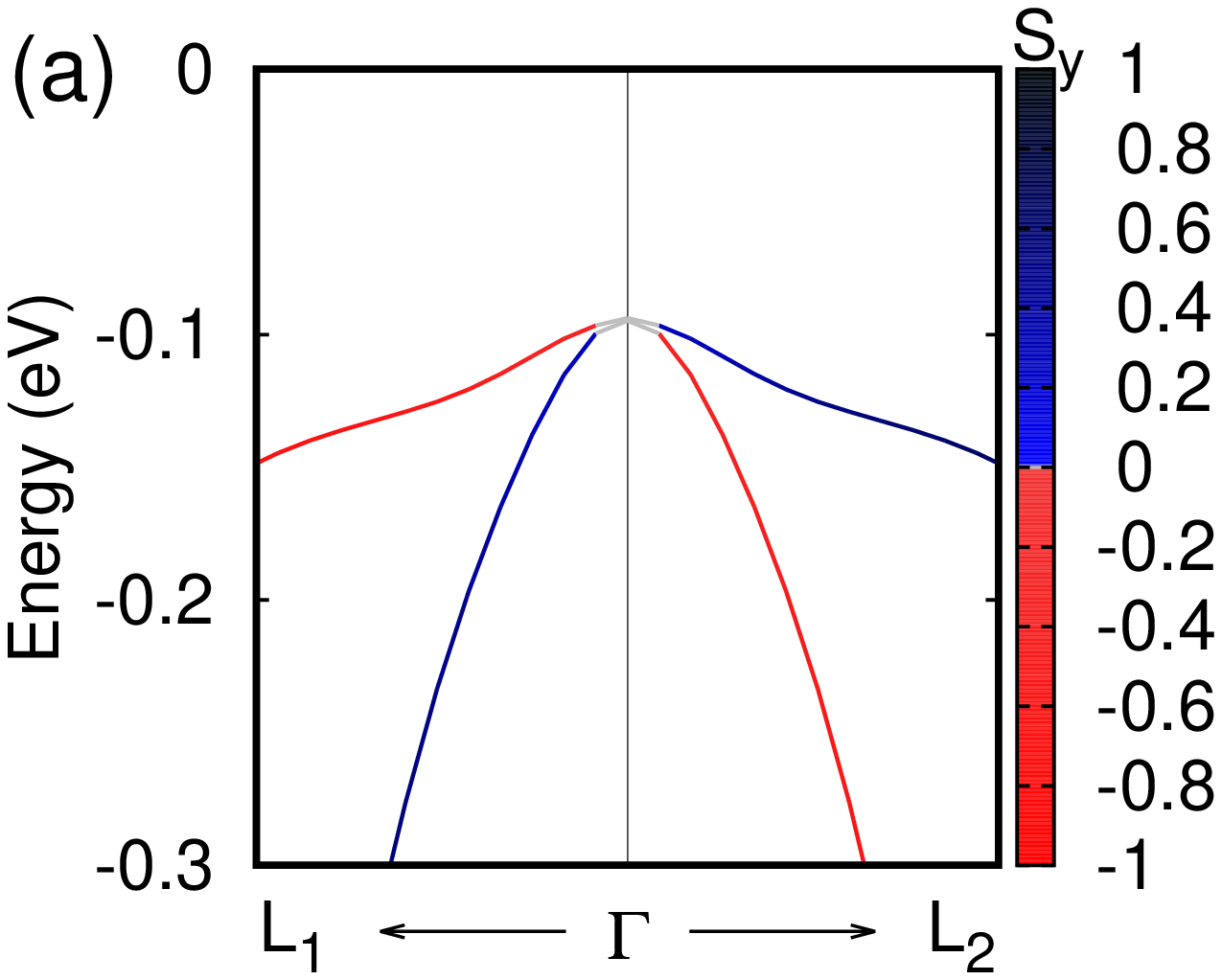}
\includegraphics[width=2.99cm,angle=270]{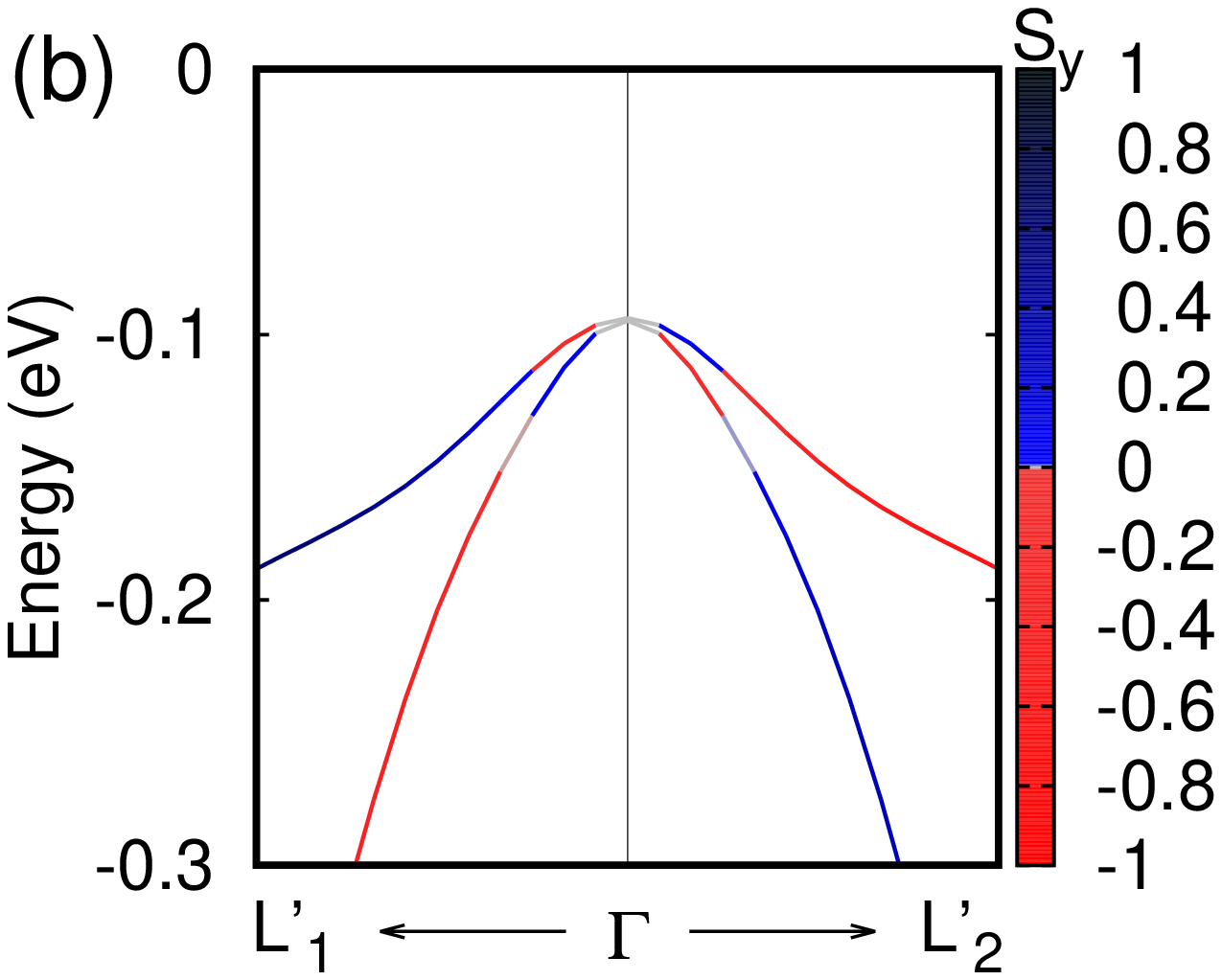}
\includegraphics[width=2.99cm,angle=270]{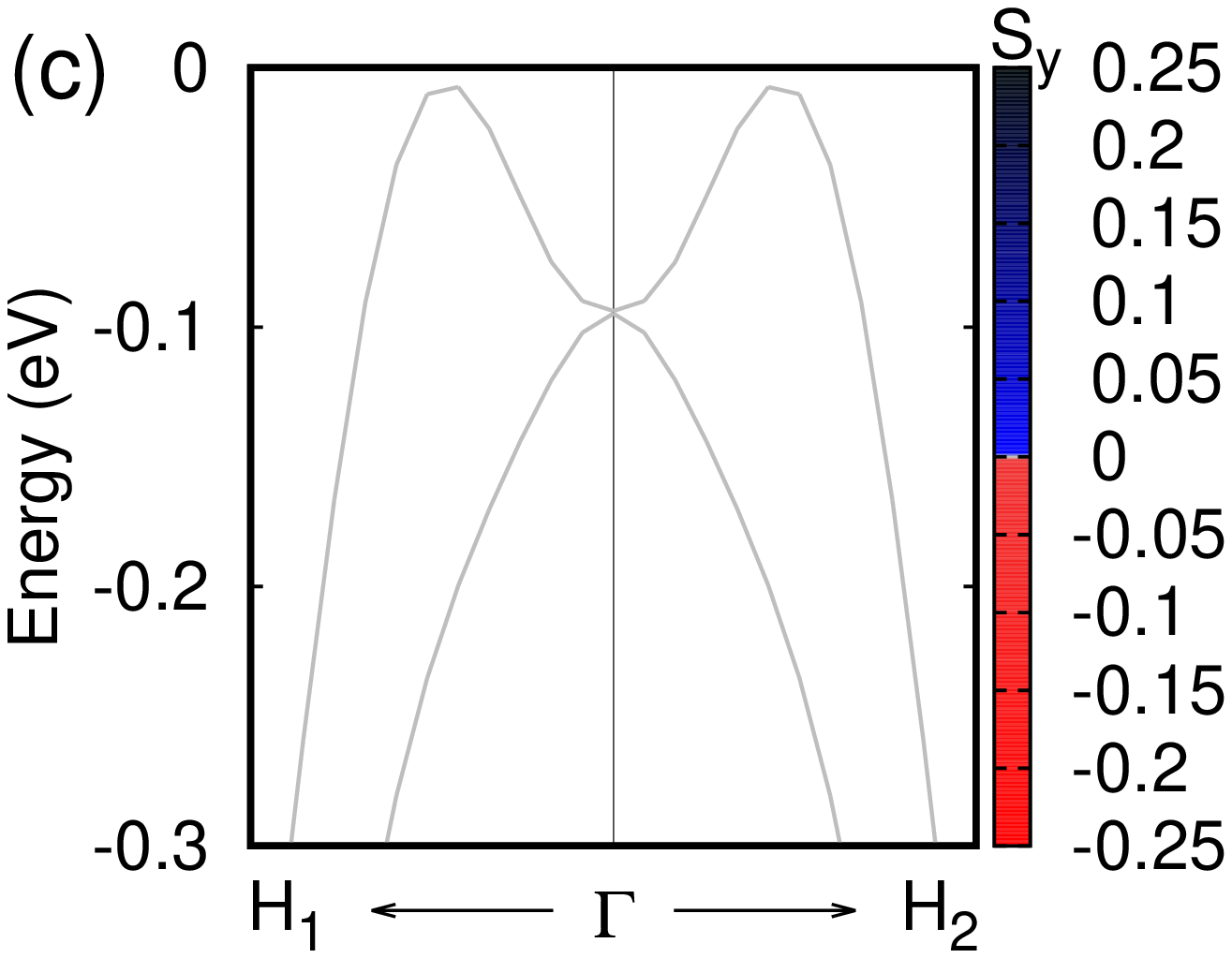}
\includegraphics[width=2.99cm,angle=270]{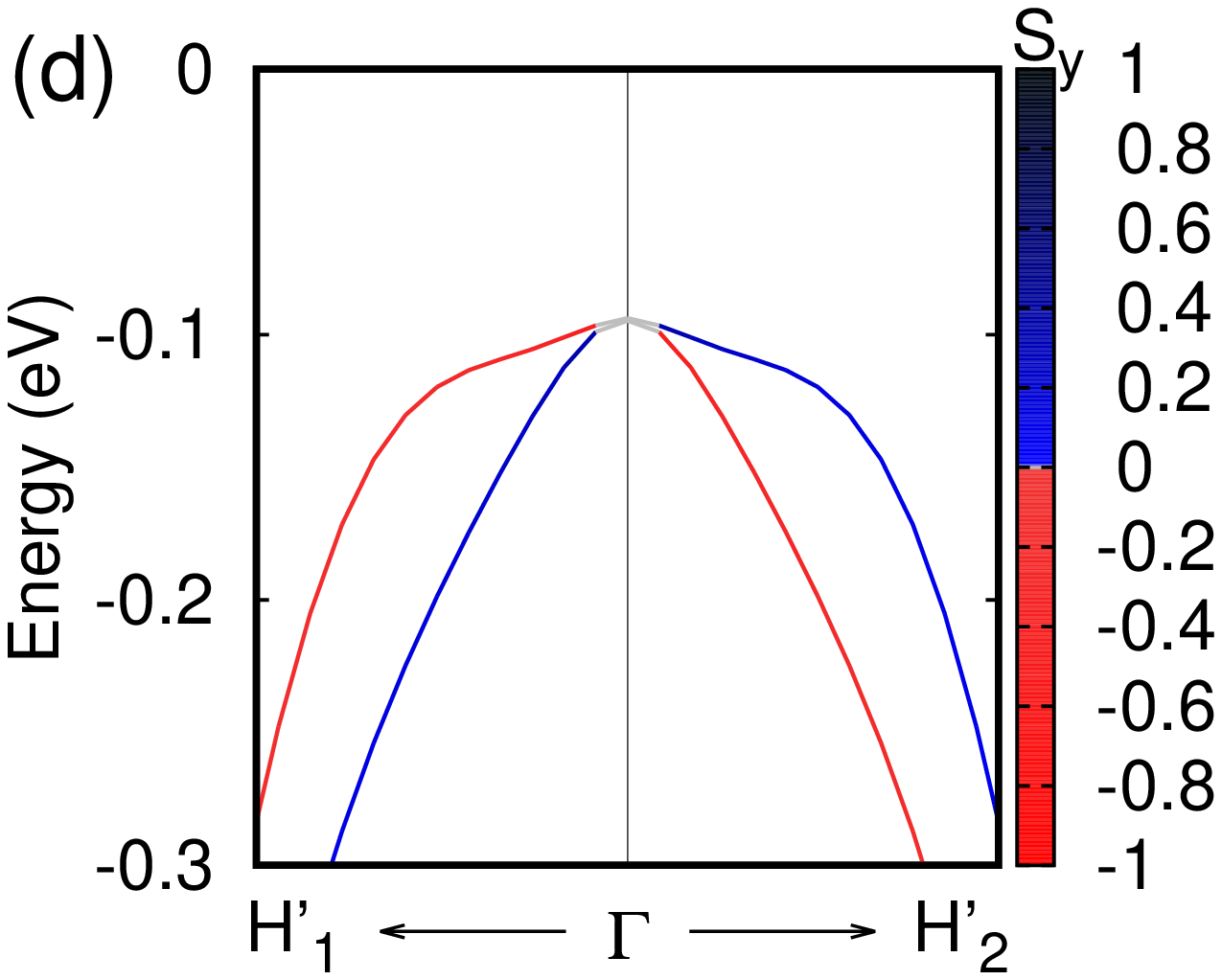}
\caption{Spin-resolved band structure of MnTe with \textbf{n} $||$ $y$-axis for the S$_y$ components along the inequivalent directions (a) L$_1$-$\Gamma$-L$_2$, (b) L'$_1$-$\Gamma$-L'$_2$, (c) H$_1$-$\Gamma$-H$_2$ and (d) H'$_1$-$\Gamma$-H'$_2$. The position of these k-points in the Brillouin zone is reported in Figure \ref{RSML_MnTe} of the main text. The plots focus on the top of the valence band from -0.3 eV up to the Fermi level. The Fermi level is set to zero.}\label{BS_Sy}
\end{figure} 


The size of the S$_z$ component, which is the dominant component, is approximately 180 times larger than the subdominant S$_x$ and S$_y$ components (see Table I of supplementary materials). However, the ratio between the sizes of the spins does not reflect in the spectral weight in the band structure. Indeed, even if the S$_x$ and S$_y$ subdominant components have spectral weight smaller than the S$_z$ component, they are still of the same order of magnitude. Therefore, the spin–momentum locking of the spin components S$_x$ and S$_y$ is a robust feature. We demonstrate that the relativistic spin-momentum locking is a fundamental property even in systems such as vanadates, where the spin–orbit coupling is intrinsically weak.

\begin{figure}[t!]
\centering
\includegraphics[width=2.99cm,angle=270]{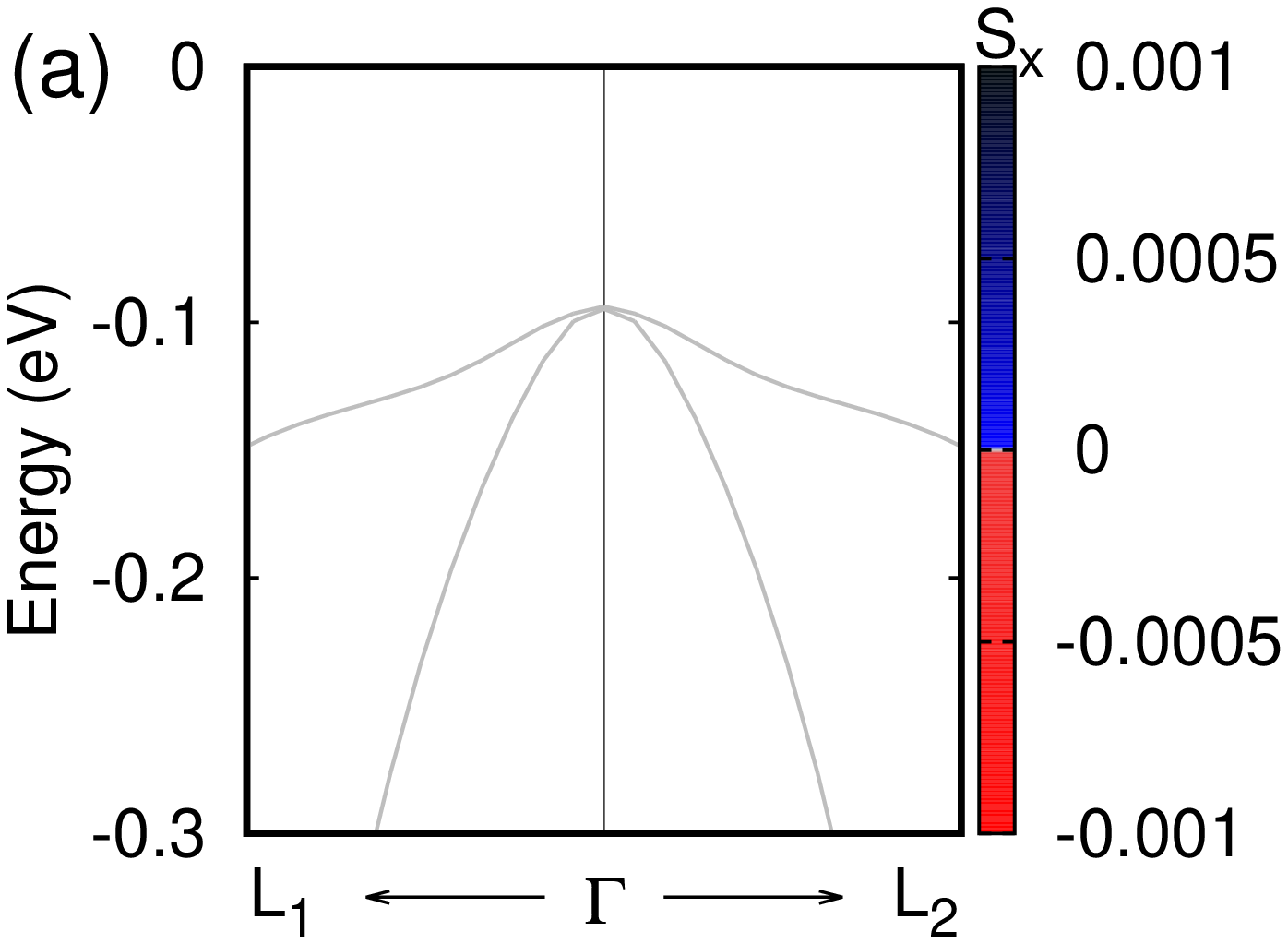}
\includegraphics[width=2.99cm,angle=270]{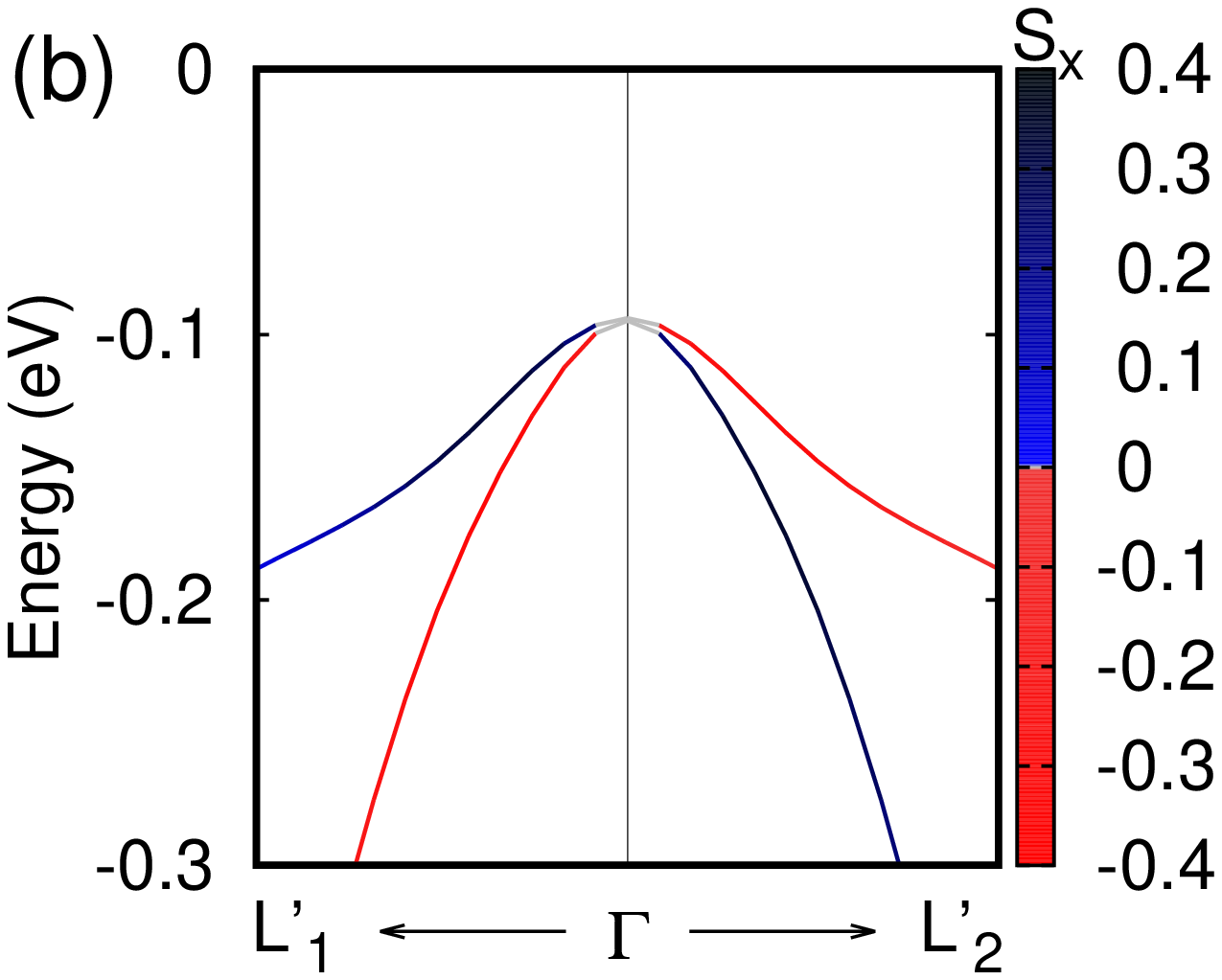}
\includegraphics[width=2.99cm,angle=270]{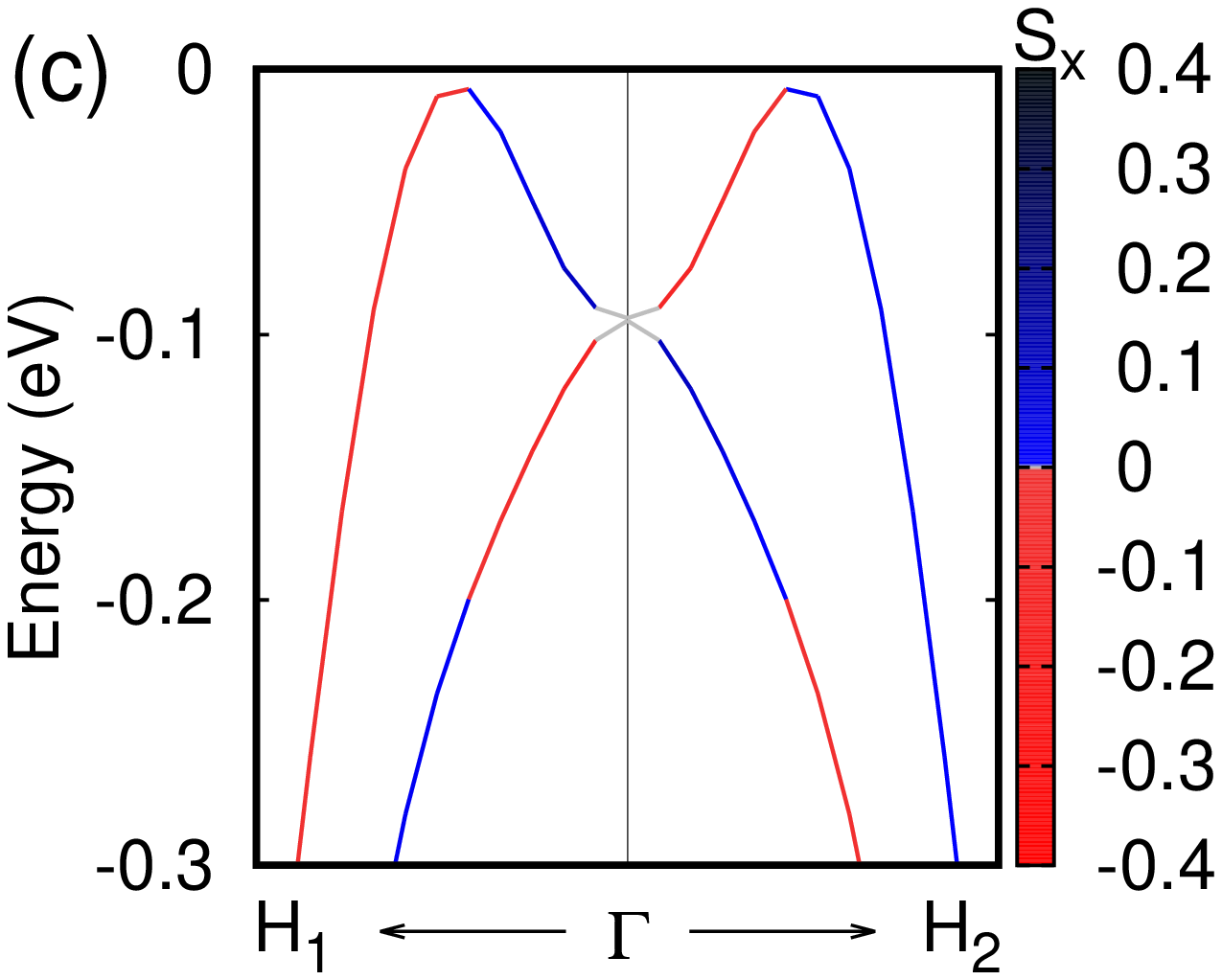}
\includegraphics[width=2.99cm,angle=270]{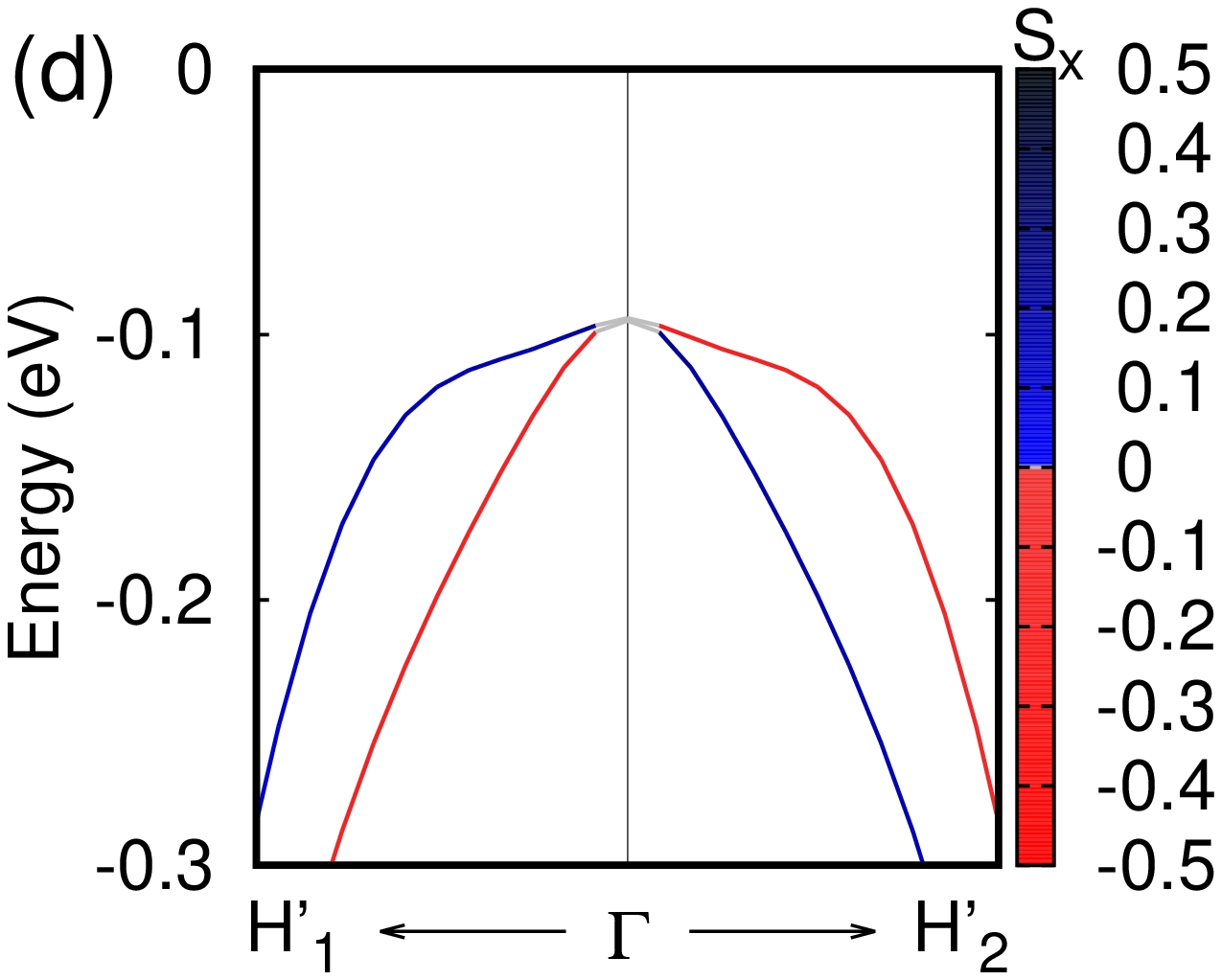}
\caption{Spin-resolved band structure of MnTe with \textbf{n} $||$ $y$-axis for the S$_x$ components along the inequivalent directions (a) L$_1$-$\Gamma$-L$_2$, (b) L'$_1$-$\Gamma$-L'$_2$, (c) H$_1$-$\Gamma$-H$_2$ and (d) H'$_1$-$\Gamma$-H'$_2$. The position of these k-points in the Brillouin zone is reported in Figure \ref{RSML_MnTe} of the main text. The plots focus on the top of the valence band from -0.3 eV up to the Fermi level. The Fermi level is set to zero.}\label{BS_Sx}
\end{figure} 

\begin{figure*}[t!]
\centering
\includegraphics[width=15.24cm]{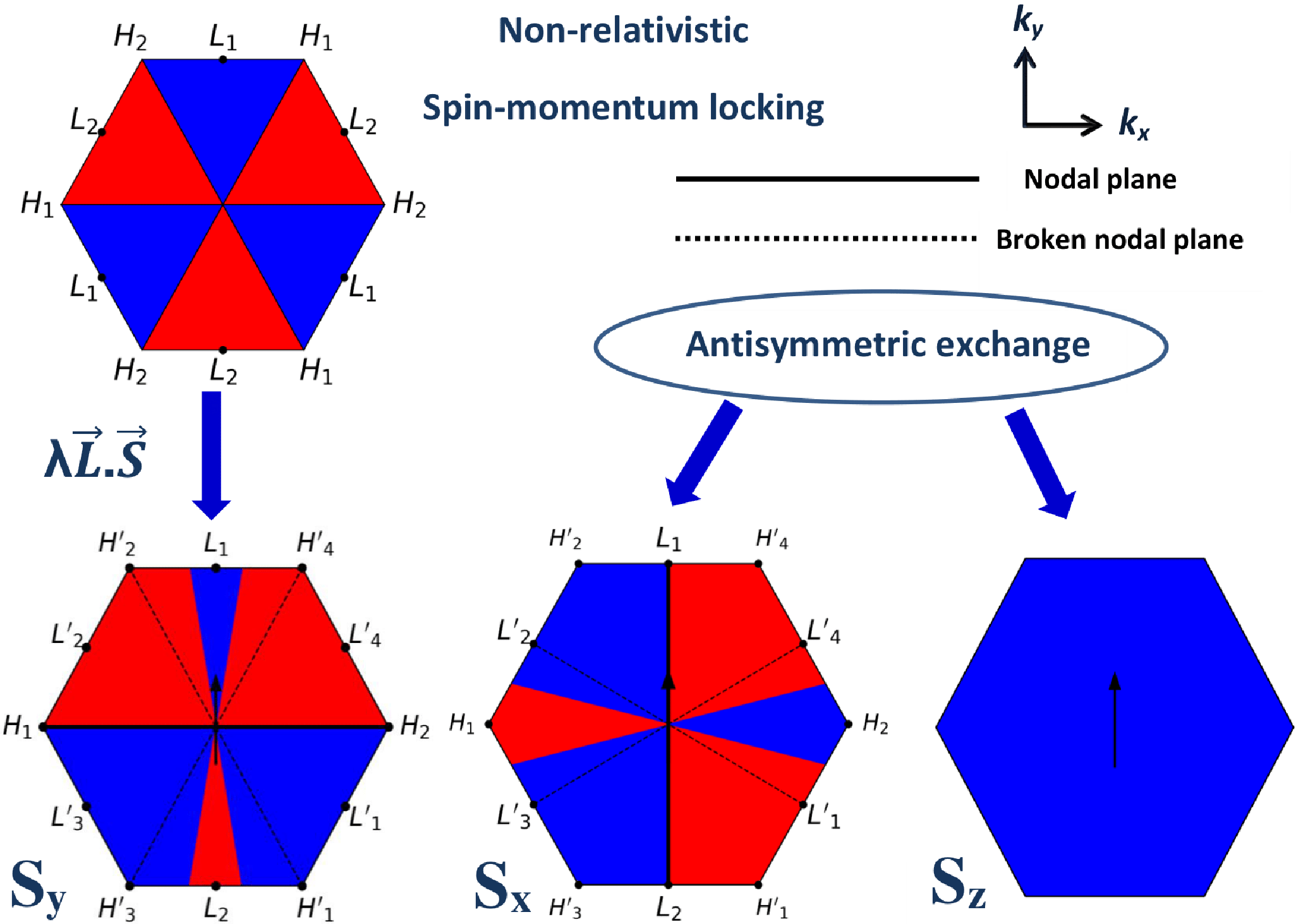}
\caption{Non-relativistic g-wave spin-momentum locking of MnTe in the top part. Relativistic spin-momentum locking for MnTe with N\'eel vector along the $y$-axis in the bottom part. The S$_y$ is the dominant component and inherits the non-relativistic spin-momentum locking, but the N\'eel vector lowers the symmetry from g-wave to d$_{yz}$-wave. The S$_z$ component is s-wave due to the weak ferromagnetism, while the spin-momentum locking of the S$_x$ component is d$_{xz}$-wave, which is rotated by $\frac{\pi}{2}$ with respect to the S$_y$ component. The solid lines represent the nodal planes, while the dashed lines represent the nodal lines broken by the N\'eel vector. Red and blue represent regions of the Brillouin zone with opposite spin-splitting. The weak ferromagnetism is represented by the s-wave with a complete Brillouin zone. The arrows in the middle of the hexagonal Brillouin zone for the relativistic case represent the N\'eel vector along the $y$-axis.}\label{RSML_MnTe}
\end{figure*} 


In the case of MnTe, the non-relativistic spin-momentum locking is composed of the bulk g-wave\cite{Krempask2024,Dzian_2025,PhysRevB.107.L100417,Mazin23}. We consider the experimental case where the N\'eel vector is along the $y$-axis\cite{PhysRevB.96.214418}. The S$_z$ component exhibits weak ferromagnetism\cite{kluczyk2023coexistence,PhysRevLett.130.036702} and it is described by the s-wave spin-momentum locking. Interestingly, the main contribution to the S$_z$ component comes from the Te atoms hosting SOC rather than Mn atoms (see Fig. S6). The S$_y$ component is the dominant component and it inherits the bulk g-wave from 
the non-relativistic spin-momentum locking protected by the C$_{6z}$ symmetry. However, the N\'eel vector is in the ab plane and breaks the C$_{6z}$ symmetry and reduces the symmetries for the spin-momentum locking of the in-plane component. We perform band structure calculations resolved for the S$_y$ and S$_x$ components reported in Figs. \ref{BS_Sy} and \ref{BS_Sx}, respectively. 
In the non-relativistic case, MnTe hosts nodal planes for the plane k$_z$=0, and the other three nodal planes orthogonal to the first nodal plane a, containing the $\Gamma$-H directions, while the maximum of the non-relativistic spin-splitting is along the $\Gamma$-L direction. 
After the breaking of the C$_{6z}$ symmetry by the N\'eel vector, we have six inequivalent L points and six inequivalent H points in the relativistic limit; these k-points are schematically reported in Fig. \ref{RSML_MnTe}. As expected, along  L$_1$-$\Gamma$-L$_2$ and  L'$_1$-$\Gamma$-L'$_2$, the signature of the spin-momentum locking for the S$_y$ component is present as displayed in Fig. \ref{BS_Sy}(a,b). From the resolved band structure of the S$_y$ component in Fig. \ref{BS_Sy}, we observe that the nodal plane persists along $\Gamma$-H, as reported in Fig. \ref{BS_Sy}(c), but disappears along $\Gamma$-H' as shown in Fig. \ref{BS_Sy}(d) where the signature of the spin-momentum locking appears. Therefore, the symmetry of the spin-momentum locking for the S$_y$ component gets reduced from bulk g-wave to d$_{yz}$-wave due to breaking of the C$_{6z}$ symmetry from the N\'eel vector. When we performed the resolved band structure calculation for S$_x$, we obtained a 
striking result: we have a nodal plane for the relativistic spin-splitting $\Gamma$-L as observed in Fig. \ref{BS_Sx}(a), which results in a d$_{xz}$-wave spin-momentum locking, which is rotated by $\frac{\pi}{2}$ with respect to the spin-momentum locking of S$_y$. The signature of the spin-momentum locking is displayed along all other directions in Fig. \ref{BS_Sx}(b,c,d).
As demonstrated in the literature\cite{Smejkal22beyond}, the non-relativistic spin-momentum locking depends on the symmetry of the spin-polarized charge density in real space. We can assume that the charge density of S$_x$ is rotated by $\frac{\pi}{2}$ with respect to the charge density of S$_y$, and this will result in the spin-momentum locking obtained numerically through DFT. 
The spin-momentum locking of S$_x$ is confirmed by the spin-resolved Fermi surface for hole-doped MnTe reported by Ye et al.\cite{ye2025dominantorbitalmagnetizationprototypical}. The spin-resolved band structure for the S$_z$ component is reported in Fig. S7 and it is consistent with the presented picture.

We can focus on the size of the spin-projection in the k-space. For the dominant component Sy, the size of spin projection in the Figures \ref{BS_Sy}(a,b,d) (the ones not involving the nodal plane) is between -1 and 1. For the subdominant component Sx, the size of spin projection in the Figures \ref{BS_Sy}(b,c,d) (the ones not involving the nodal plane) is bounded to be less than 0.4-0.5. Therefore, even if in the real space there are more than 3 orders of magnitude of difference, in the k-space the spin-projections are of the same order.
The small spin expectation in the real space results from the cancellations between contributions at
different momenta.

The lowering of the symmetry from g-wave to d-wave under structural perturbation was reported in recent papers\cite{ssxp-gz9l,karetta2025straincontrolledgdwave} in the non-relativistic limit. Even if the spin-momentum locking resembles the original g-wave in the case of a small perturbation, the spin-momentum locking is mathematically a d-wave since the g-wave nodal planes are broken. In MnTe, the breaking of the g-wave symmetry is of relativistic origin. The relativistic spin-momentum locking of MnTe with N\'eel vector along the $y$-axis is composed of d$_{xz}$-wave, d$_{yz}$-wave and s-wave for the S$_x$, S$_y$ and S$_z$ components, respectively. The top view of the relativistic spin-momentum locking of MnTe is schematically reported in Fig. \ref{RSML_MnTe}.


While relativistic spin-momentum locking is intrinsically present in the DFT simulations, our results suggest a revisit of the model Hamiltonian studies. We can write the contributions to the non-relativistic Hamiltonian for an effective single-orbital in terms of Pauli matrices for spin and site in the following Hamiltonian:
\begin{align} 
\label{eq:H0skxkz}
\mathcal{H}^0  =& \, \, \, \varepsilon(\boldsymbol{k})\sigma_0^{spin}\sigma_0^{site} \\ 
\nonumber 
\mathcal{H}^{\rm AM}_{Sz}  = & \Delta_z\sigma_z^{spin}\sigma_z^{site} \\ 
\label{eq:HAMskxkz} & +4t_{am}\sin{k_x}\sin{k_z}\sigma_0^{spin}\sigma_z^{site}  
\end{align}
where $\sin{k_x}\sin{k_z}$ is the spin-momentum locking of the S$_z$ component of YVO$_3$.
For the two subdominant components, the hopping producing the spin-momentum locking is activated by the spin-orbit coupling $\lambda$ via the antisymmetric exchange. We named $\Delta_x$ and $\Delta_y$ the spin-splitting for the relative component and the equations to be:
\begin{equation}\label{H_Sx}
\mathcal{H}^{\rm AM}_{Sx}=\Delta_x\sin(k_x)\sin(k_z)\sigma_y^{spin}\sigma_z^{site}
\end{equation}
and 
\begin{equation}
\mathcal{H}^{\rm AM}_{Sy}=\Delta_y\sin(k_y)\sin(k_z)\sigma_y^{spin}\sigma_z^{site}
\end{equation}
Additional information can be found in the Supplementary Materials. 
This model Hamiltonian should be suitable for implementing electronic correlations or studying the topological properties in altermagnets once $\varepsilon(\boldsymbol{k})$ exhibits a band inversion.


In all cases analyzed in the main text and supplementary materials, we did not find identical spin–momentum locking for two different spin components. In all investigated cases, the spin-momentum lockings of the spin-components have symmetries equal to or lower than those of the non-relativistic spin–momentum locking. 
Other centrosymmetric compounds that are hosts of relativistic spin–momentum locking in all three components include Ca$_2$RuO$_4$\cite{Amar214},   Co$_{0.25}$NbSe$_2$\cite{devita2025opticalswitchinglayeredaltermagnet,gong2025tunabilitymagneticpropertiesni} and several others. Such locking may also play a role in systems with broken PT symmetry that do not have a collinear counterpart, as exemplified by kagome lattices.\cite{Cheong2024,10.1063/5.0283630}.
The scenario changes once inversion symmetry is broken, giving rise to relativistic band features such as the Rashba effect\cite{leon2025hybriddpwavealtermagnetismca3ru2o7,Wang2025,Zhang2025} or the persistent spin helix\cite{tenzin2025persistentspintexturesaltermagnetism}. 
Beyond the even-wave spin–momentum locking, the resulting state acquires a relativistic $p$-wave character, forming a hybrid system with mixed $p$- and $d$-wave contributions of unequal weight\cite{leon2025hybriddpwavealtermagnetismca3ru2o7}, resulting in a loss of the wave symmetry for the spin-momentum locking. Nevertheless, a component of the spin-momentum locking can survive in the relativistic limit. For example, suppose inversion symmetry is broken by an electric field along the $z$-axis. In that case, the Rashba effect will involve the S$_x$ and S$_y$ components, leaving intact the spin–momentum locking of the spin component parallel to the electric field.\cite{Amar214}.

Our results call for further investigation into the antisymmetric exchange in altermagnets. While Roig et al. calculated the leading antisymmetric term responsible for weak ferromagnetism\cite{839n-rckn}, additional terms lead to spin canting with zero net magnetization. Although the spin cantings are extremely small compared to the spin magnitude, they significantly modify the spectral weight in the k-space. Incorporating these terms into theoretical models is therefore essential for studying relativistic spin–momentum locking within the framework of model Hamiltonians. Such extensions may be crucial for understanding the interplay between topology and magnetism, whereas previous studies of altermagnets have primarily focused on systems with single-component momentum locking\cite{PhysRevLett.134.096703,Li2025,PhysRevB.110.125129}, our results point to new challenges in exploring topological phenomena in the presence of relativistic spin–momentum lockings.

The experimental verification of relativistic spin–momentum locking will require high-energy spin-resolved ARPES capable of probing bulk states in proposed centrosymmetric altermagnet candidates\cite{Krempask2024}.
Our work establishes relativistic spin–momentum locking as a comprehensive framework for characterizing altermagnets. 
While the canted spin components in real space are three orders of magnitude smaller than the dominant component, their contribution in $\mathbf{k}$-space is of comparable magnitude. Consequently, they can strongly influence spin-related physical properties evaluated in $\mathbf{k}$-space, such as the spin Hall conductivity~\cite{hirakida2025multipoleanalysisspincurrents} or the spin photocurrent.
Our work offers the framework for further studies using density functional theory on multipoles\cite{PhysRevX.14.011019,ko2025magneticoctupolehalleffect,lxkx-ypbg,PhysRevX.15.031006,doi:10.7566/JPSJ.93.072001} and quantum metrics in altermagnets\cite{PhysRevLett.133.106701} including all three components. Beyond the dominant components, we demonstrate that altermagnetic spintronic devices can also harness the spin–momentum locking of the canted components. Moreover, these insights indicate that device architectures must take into account that the even-wave symmetry of altermagnets can be reduced as they evolve from the non-relativistic to the relativistic limit.

\section*{Acknowledgments}

The authors thank  J. Skolimowski, T. Dietl, V. V. Volobuev, G. Cuono, C. C. Ye and J. S{\l}awi{\'n}ska for useful discussions.  
This research was supported by the "MagTop" project (FENG.02.01-IP.05-0028/23) carried out within the "International Research
Agendas" programme of the Foundation for Polish Science, co-financed by the
European Union under the European Funds for Smart Economy 2021-2027 (FENG). C.A. acknowledges support from PNRR MUR project PE0000023-NQSTI.
We further acknowledge access to the computing facilities of the Interdisciplinary Center of Modeling at the University of Warsaw, Grant g91-1418, g91-1419, g96-1808 and g96-1809 for the availability of high-performance computing resources and support. We acknowledge the access to the computing facilities of the Poznan Supercomputing and Networking Center, Grants No. pl0267-01, pl0365-01 and pl0471-01.

\bibliography{references}
\end{document}